\documentclass[preprint,prd,noshowpacs,nofootinbib]{revtex4}

\usepackage{color} 
\usepackage{amssymb}
\usepackage{amsmath}
\usepackage{graphicx} 
\usepackage{float}
\usepackage{braket}     
\usepackage{pdfpages}
\usepackage{epstopdf}
\usepackage{appendix}
\usepackage[utf8]{inputenc}

\begin{document}

\title{Electromagnetic Field Momentum in Theoretical Magnetic Monopole-like Models}

\author{Isaiah Ruz}
\email{iruz@mail.fresnostate.edu}
\affiliation{Physics Department, California State University Fresno, Fresno, CA 93740 \\}

\author{Douglas Singleton}
\email{dougs@mail.fresnostate.edu}
\affiliation{Physics Department, California State University Fresno, Fresno, CA 93740 \\}

\date{\today}

\begin{abstract}
We show that some standard models of magnetic charge --  the Banderet potential, the Dirac string potential, and the Wu-Yang fiber bundle approach -- carry an electromagnetic field momentum in the presence of an electrostatic field. This electromagnetic field momentum suggests a violation of the center-of-energy theorem. We show that each of these monopole models also carry a hidden mechanical momentum, which saves the center-of-energy theorem. 
\end{abstract}

\maketitle

\section{Introduction}

In electromagnetism the electric and magnetic fields can carry both momentum and angular momentum. Using the notation and units of \cite{jackson} the field momentum and field angular momentum are given by 
\begin{equation}
    \label{momentum}
    {\bf P}_{{\rm EM}} = \frac{1}{4 \pi } \int ({\bf E} \times {\bf B}) d^3 x ~,
\end{equation}
and
\begin{equation}
    \label{ang-momentum}
    {\bf L}_{{\rm EM}} = \frac{1}{4 \pi } \int  {\bf x} \times ( {\bf E} \times {\bf B}) d^3 x ~.
\end{equation}
We have set the speed of light to unity ($c=1$) throughout the paper. Otherwise, the pre-factor for both equations would be $\frac{1}{4 \pi c}$. 
Including these contributions to momentum and angular momentum is essential when applying the conservation laws to systems in which momentum or angular momentum can be exchanged between particles and fields.

In reference \cite{shockley} Shockley and James considered a magnetic dipole composed of two counter-rotating electrically charged disks and two opposite electric charges. As the disks are brought to rest, there is an apparent violation of momentum conservation. Shockley and James proposed that there must be some momentum hidden in the rotating disks to preserve momentum conservation. Subsequently, Coleman and Van Vleck \cite{coleman} showed that there was indeed hidden momentum in the system addressing the paradox raised in \cite{shockley}.

Hidden momentum is a well-known concept that has been discussed in various articles (see the resource letter \cite{griffiths-re} for an extensive list) and textbooks \cite{griffiths-text,jackson-3}. In this paper, we show that hidden mechanical momentum plays a role in three different models of magnetic charge. 

\section{Monopole Models}

Including magnetic charge into Maxwell's equations is straightforward. In terms of electric and magnetic fields and electric and magnetic sources, the extended Maxwell equations are given by 
\begin{eqnarray}
\label{e-maxwell3-source1}
&&\nabla \cdot {\bf E}  = 4 \pi \rho _{e}  ~~~;~~~ \nabla \times {\bf E} + \frac{\partial {\bf B}}{\partial t} =  - 4 \pi {\bf J}_{m}  \\ 
\label{b-maxwell3-source1}
&& \nabla \cdot {\bf B} =  4 \pi \rho _{m} ~~~;~~~ \nabla \times {\bf B} - \frac{\partial {\bf E}}{\partial t} = 4 \pi {\bf J}_{e} ~,
\end{eqnarray}
where the $e$ and $m$ subscripts indicate electric and magnetic charge and current densities. In the standard Maxwell equations $\rho _m=0$ and ${\bf J}_m=0$. 

The formulation of Maxwell's equations in terms of potentials can be extended to included magnetic monopoles, but it requires {\it two} scalar potentials and {\it two} vector potentials:   
\begin{equation}
    \label{dual-potential}
    {\bf E} = - \nabla \phi_e - \frac{\partial{\bf A}}{\partial t} - \nabla \times {\bf C}~~~; ~~~ {\bf B} = -\nabla \phi_m - \frac{\partial{\bf C}}{\partial t} + \nabla \times {\bf A} ~.
\end{equation}
This is the Cabibbo-Ferrari approach to magnetic charge \cite{cabibbo}. Two methods for dealing with the second scalar potential, $\phi_m$, and the second vector potential, ${\bf C}$ are discussed in \cite{zwanziger,singleton-1996}. 

The models considered in this paper lie within ordinary classical electrodynamics; they do not include free magnetic monopole sources or involve doubling of the potentials, but rather are in the context of the standard single scalar and single vector potential 
\begin{equation}
    \label{potential}
    {\bf E} = - \nabla \phi_e - \frac{\partial{\bf A}}{\partial t} ~~~; ~~~ {\bf B} = \nabla \times {\bf A} ~,
\end{equation}
so that one has  $\nabla \cdot {\bf B} = \nabla \cdot (\nabla \times {\bf A}) = 0$.
In the next three subsections we will discuss how Dirac and others dealt with this zero divergence of the magnetic field in the case where the potentials are given by \eqref{potential}. In all three cases, the zero divergence of ${\bf B}$ is achieved by balancing the outward flux of the magnetic Coulomb field with some compensating inward flux. 

\subsection{Dirac String Potential}

The oldest and best known model for a monopole-like field within standard electrodynamics is due to Dirac \cite{dirac,dirac1} (several good reviews can be found in \cite{shnir,heras,olive,blag,felsager,mavromatos}); it is based on the string potential:
\begin{equation}
\label{A-coulomb}
    {\bf A} _\pm ({\bf x}) = \frac{g}{r} \left( \frac{\pm 1 - \cos \theta }{\sin \theta} \right) {\bf{\hat \varphi}} ~.
\end{equation}
Equation \eqref{A-coulomb} shows two versions of the Dirac vector potential, ${\bf A} _+ ({\bf x})$ and ${\bf A} _- ({\bf x}) $. Both vector potentials have a linear string singularity: ${\bf A} _+ ({\bf x})$ is singular along the $-z$ axis ({\it i.e.} $\theta =\pi$) but is regular along the $+$z axis,  and ${\bf A} _- ({\bf x})$ is singular all along the $+z$ axis ({\it i.e.} $\theta =0$), but regular along the $-$ z axis. The two vector potentials in \eqref{A-coulomb} are related by the gauge transformation ${\bf A} _+ ({\bf x}) = {\bf A} _- ({\bf x}) + \nabla  \alpha$ with the non-single-valued gauge function $\alpha = 2 g \varphi$, which yields $\nabla \alpha = (2g)/(r \sin \theta) {\hat \varphi}$. Except on the z-axis, the curl of the vector potentials in \eqref{A-coulomb}, gives a Coulomb magnetic field ${\bf B}^{{\rm Coulomb}} = \nabla \times {\bf A} _\pm ({\bf x}) = g\frac{{\hat {\bf r}}}{r^2}$.

Taking a closer look at the string singularities in ${\bf A} _+ ({\bf x})$  and ${\bf A} _- ({\bf x})$, one can show that each potential has the physical interpretation of an infinitesimally thin solenoid which carries an inward magnetic flux of $-4 \pi g$. This inward flux then balances the outward magnetic flux of $4 \pi g$ coming from the Coulomb magnetic field. Note that a monopole potential can also be formed by averaging the two potentials in \eqref{A-coulomb} as $\frac{1}{2} ( {\bf A}_+ + {\bf A}_-) = \frac{g}{r} \frac{\cos \theta}{\sin \theta} {\hat \varphi}$. This average configuration has string singularities along both the positive and negative $z$ axes, with an inward flux of $-2 \pi g$ along each string singularity for a total of $-4 \pi g$. To see this interpretation of the string singularities in \eqref{A-coulomb} as an infinitesimally thin solenoid, we rewrite the vector potentials in cylindrical coordinates which are more easily applied to the cylindrical symmetry of a solenoid. This turns the expression in \eqref{A-coulomb} into 
\begin{equation}
    \label{A-coulomb1}
    {\bf A} _\pm ({\bf x}) =\frac{g}{\rho} \left( \pm 1 - \frac{z}{\sqrt{\rho^2+z^2}} \right) {\bf{\hat \varphi}}~.
\end{equation}
Next we regularize the string singularities of \eqref{A-coulomb1} as \footnote{The details of the calculation starting with \eqref{A-coulomb2} and leading to the final result in \eqref{B-coulomb} are straightforward, but lengthy. The full details of this calculation can be found in Appendix D of reference \cite{heras}.}
\begin{equation}
    \label{A-coulomb2}
    {\bf A}^\varepsilon _\pm ({\bf x}) =\frac{g \Theta(\rho - \varepsilon)}{\rho} \left( \pm 1 - \frac{z}{\sqrt{\rho^2+z^2 + \varepsilon ^2}} \right) {\bf{\hat \varphi}}~,
\end{equation}
where $\Theta(x) $ is a step function ({\it i.e.} $\Theta (x) =0$ if $x<0$ and $\Theta (x) =1$ if $x \ge 0$) and $\varepsilon$ is a small quantity which can be taken to zero. The regularized vector potential in \eqref{A-coulomb2} excises the string singularity. Taking the curl of ${\bf A}^\varepsilon _\pm ({\bf x})$ and letting $\varepsilon \to 0$ gives the magnetic field
\begin{eqnarray}
    \label{B-coulomb}
    {\bf B} &=& \lim_{\varepsilon \to 0} (\nabla \times {\bf A}^\varepsilon _\pm ) = g \frac{\rho {\hat {\bf \rho}} + z {\hat {\bf z}}}{(\rho ^2 +z^2)^{3/2}}\pm 2  g \frac{\delta (\rho)}{\rho} \Theta (\mp z) {\bf {\hat z}} 
   \nonumber \\ 
   &=& g \frac{{\hat {\bf r}}}{r^2} \pm 4 \pi g \delta (x) \delta (y) \Theta (\mp z) {\hat {\bf z}} ~.
\end{eqnarray}
In the last step, we have converted back from cylindrical coordinates to spherical coordinates for the Coulomb term and to Cartesian coordinates for the delta function term. 

The Coulomb term in \eqref{B-coulomb} gives an outward magnetic flux of $4 \pi g$, while the delta function term in \eqref{B-coulomb} gives an inward magnetic flux of $-4 \pi g$. Thus, the magnetic field in \eqref{B-coulomb} has zero net flux since the outward Coulomb flux is balanced by the inward solenoidal flux. The details of this calculation are given in Appendix B.  

The result for the magnetic field in \eqref{B-coulomb} can be found in several review articles and monographs \cite{shnir,heras,olive,blag,felsager,adorno,mavromatos}. To make the magnetic field in \eqref{B-coulomb} appear as a Coulomb magnetic field, the solenodial part of the magnetic field is ``hidden" from any electric charge, $q,$ that is placed in the vicinity of the magnetic charge, $g$. This hiding of the string part of the magnetic field in \eqref{B-coulomb} is accomplished by imposing the Dirac quantization condition \cite{dirac,dirac1,shnir,heras,olive,blag,felsager,adorno,mavromatos}
\begin{equation}
    \label{dirac-cond}
    qg = n \frac{\hbar}{2} ~,
\end{equation}
{\it i.e.} the product of the charges $q$ and $g$ is equal to an integer ({\it i.e.} $n$) multiple of Planck's reduced constant divided by 2. There are many ways to obtain the famous quantization condition in \eqref{dirac-cond} that can be found in various textbooks, review articles, and monographs \cite{jackson,shnir,heras,olive,blag,felsager,adorno,mavromatos}. 
For example, \eqref{dirac-cond} can be derived by requiring that the Aharonov-Bohm phase \cite{aharonov} be undetectable {\it i.e.} that 
\begin{eqnarray}
\label{ab-condition}
\exp \left[- \left( \frac{iq}{\hbar} \right) \oint {\bf A}_{string} \cdot d {\bf l} \right] &=& \exp \left[- \left( \frac{iq}{\hbar} \right) \int_S {\bf B}_{string} \cdot d {\bf a} \right] \nonumber \\
&=& \exp \left[\left( \frac{iq}{\hbar} \right) 4 \pi g \right]=1 \to \frac{4 \pi qg}{\hbar} = 2 \pi n
\end{eqnarray}
which leads to \eqref{dirac-cond}. 

Next, we calculate the current density associated with the magnetic field in \eqref{B-coulomb} via ${\bf J}_e = \frac{\nabla \times {\bf B}}{4 \pi}$. The curl of the Coulomb part of \eqref{B-coulomb} gives zero while the curl of the solenoidal string part gives 
\begin{eqnarray}
    \label{j}
    {\bf J}_e &=& \nabla \times \left( \pm g \delta (x) \delta (y) \Theta (\mp z) {\hat {\bf z}}\right) \nonumber \\
    &=& \pm g \Theta (\mp z) \left[ \delta (x) \delta ' (y) {\bf \hat x} - \delta (y) \delta' (x) {\bf \hat y} \right] ~.
\end{eqnarray}
This result will be used later when we calculate the hidden momentum which is the negative of the volume integral of the product of the scalar potential of the electric charge $q$, and the current density ${\bf J}_e$ associated with the Dirac string {\it i.e.} ${\bf P}_{hid}= - \int \phi_e {\bf J}_e d^3x$. 

One final point is that the ${\bf B}$ field in \eqref{B-coulomb} can be physically interpreted as a line of magnetic dipoles that extends from the origin to infinity. One can write ${\bf B}$ in terms of an ${\bf H}$ field and a magnetization, ${\bf M}$, as
\begin{equation}
    \label{magnetization}
    {\bf B} = {\bf H} + 4 \pi {\bf M}~.
\end{equation}
The ${\bf H}$ field can be written as the gradient of a scalar potential ${\bf H} = - \nabla \phi_m$ (see Section 5.9 of \cite{jackson-3}) and choosing a Coulomb scalar potential, $\phi_m = \frac{g}{r}$ gives a Coulomb ${\bf H}$-field ${\bf H} = g \frac{{\hat {\bf r}}}{r^2}$. The magnetization vector provides the delta function term in \eqref{B-coulomb} with ${\bf M}  = \pm g \delta (x) \delta (y) \Theta (\mp z) {\hat {\bf z}}$. The curl of ${\bf M}$ gives the current density in \eqref{j} ${\bf J}_e = \nabla \times {\bf M} = \pm g \Theta (\mp z) \left[ \delta (x) \delta ' (y) {\bf \hat x} - \delta (y) \delta' (x) {\bf \hat y} \right]$. 

\subsection{Banderet Potential}

The Banderet approach to magnetic charge \cite{banderet} is the least well-known of the three approaches discussed in this paper. The Banderet vector potential has the form \footnote{There is a typo in \cite{shnir} which has the vector potential \eqref{band} in the ${\hat \varphi}$ direction rather than in the correct ${\hat \theta}$ direction, as can be cpnfirmed by direct calculation.}
\begin{equation}
    \label{band}
    {\bf A}_{\rm{band}} = - \frac{g}{r} \varphi \sin \theta ~ {\bf \hat \theta}~.
\end{equation}
Although this vector potential does not have a string singularity like the Dirac potential, it does have the pathology of being non-single valued due to $\varphi$. Taking the curl of ${\bf A}_{band}$ gives \cite{shnir} 
\begin{equation}
    \label{band-2}
    {\bf B} = \nabla  \times {\bf A}_{\rm{band}} = \frac{g}{r^2} {\bf \hat r} - 2 \pi g \Theta (x) \delta (y) x \frac{{\bf \hat r}}{r^2} ~.
\end{equation}
The $\frac{g}{r^2} {\bf \hat r}$ term in \eqref{band-2} is the Coulomb magnetic field of strength $g$, and the second term, $- 2 \pi g \Theta (x) \delta (y) x \frac{{\bf \hat r}}{r^2}$, is due to the non-single-valued nature of ${\bf A}_{{\rm band}}$. This second term carries an inward magnetic flux of $-4 \pi g$. This balances the outward magnetic flux of $+4 \pi g$ coming from the Coulomb term, $\frac{g}{r^2} {\bf \hat r}$, verifying that the net divergence of the ${\bf B}$-field in \eqref{band-2} is zero. This calculation of the balancing of the fluxes for the Banderet potential is carried out in Appendix B. The calculations in Appendix B show that the second term in \eqref{band-2} is needed to balance ingoing and outgoing magnetic fluxes.\footnote{We have not been able to find a more direct way to obtain this second term, nor does reference \cite{shnir} give a derivation, but simply states that \eqref{band-2} is the correct ${\bf B}$-field following from the Banderet vector potential \eqref{band}.} 

The current density associated with \eqref{band-2} can be calculated via ${\bf J}_e = \frac{\nabla \times {\bf B}}{4 \pi}$. The calculation of the current density can be found in Appendix A and gives 
\begin{eqnarray}
    \label{j2}
    {\bf J}_e =   \frac{g\Theta (x)}{2 r^3}  \left( x^2 \delta ' (y) {\bf \hat z}  -xz \delta ' (y) {\bf \hat x}  +  z \delta (y) {\bf \hat y}\right)~.
\end{eqnarray}
We will use this result later to calculate the hidden momentum carried by the Banderet potential. 

\subsection{Wu-Yang Fiber Bundle}

\begin{figure}[ht]
 \centering
\includegraphics[width=100mm]{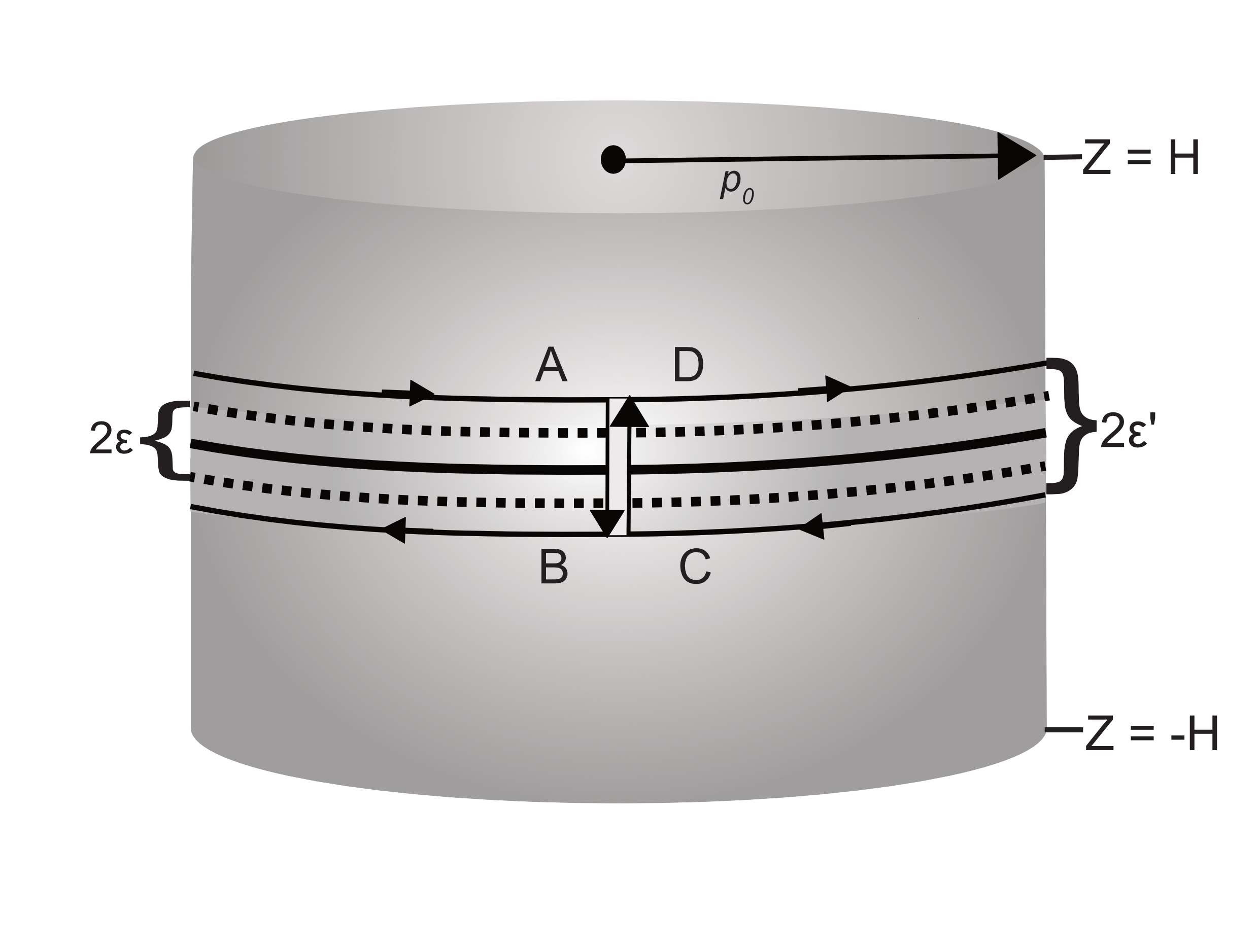}
 \caption{The line integral and cylindrical surface for evaluating $\oint {\bf A} \cdot d {\bf l}$ across the equator. This will be checked against the two possible areas in Figs. \eqref{fig2} and \eqref{fig3} for evaluating $\int {\bf B} \cdot d {\bf a}$. The thin solid line is the contour of $\oint {\bf A} \cdot d {\bf l}$, while the dotted lines denote the region of overlap between the vector potentials ${\bf A}_+$ and ${\bf A}_-$.}
\label{fig1}
\end{figure}

The Wu-Yang approach to magnetic charge \cite{wu-yang,yang} begins with the Dirac string potentials
\cite{dirac, dirac1} and applies the mathematics of fiber bundles to avoid the string singularities of \eqref{A-coulomb}. In simple terms, the vector potential over the northern hemisphere is defined as ${\bf A}_+$, while the vector potential over the southern hemisphere is defined as ${\bf A}_-$. There is a thin overlap region around the equator that has a width $2 \epsilon$ where ${\bf A}_+$ and ${\bf A}_-$ overlap. This is shown in Fig. \eqref{fig1} with a cylindrical surface instead of the usual spherical surface. Vector potentials ${\bf A}_+$ and ${\bf A}_-$ should be continuously connected across the overlap region $2 \epsilon$, for example, by a linear function. The particular form of how ${\bf A}_+$ and ${\bf A}_-$ are continuously connected is not crucially important since both ${\bf A}_+$ and ${\bf A}_-$ are in the ${\hat \varphi}$ direction, and the segments of the line integral in the overlap region are in the ${\hat {\bf z}}$ direction. Thus, ${\bf A}_\pm \cdot d{\bf l} =0$ in the overlap region and these segments do not contribute, independent of how ${\bf A}_+$ and ${\bf A}_-$ are connected across this region. 

Using boundary conditions across the equator in Fig. \eqref{fig1} we will show that the Wu-Yang construction also requires a delta function magnetic field in addition to the Coulomb magnetic field. In cylindrical coordinates our claim is that the full magnetic field is of the form
\begin{equation}
    \label{b-disk}
{\bf B} = g \frac{\rho {\hat {\bf \rho}} + z {\hat {\bf z}}}{(\rho ^2 +z^2)^{3/2}} - \frac{2g \delta (z)}{\rho} {\hat {\bf \rho}}~.
\end{equation}
The first term is the Coulomb magnetic field, and the second term is the extra delta function magnetic field. The delta function term in \eqref{b-disk} gives rise to an inward magnetic flux of $-4 \pi g$ to balance the outward flux of $4 \pi g$ due to the Coulomb term. The details of this calculation are found in Appendix B. 

\begin{figure}[ht]
 \centering
\includegraphics[width=100mm]{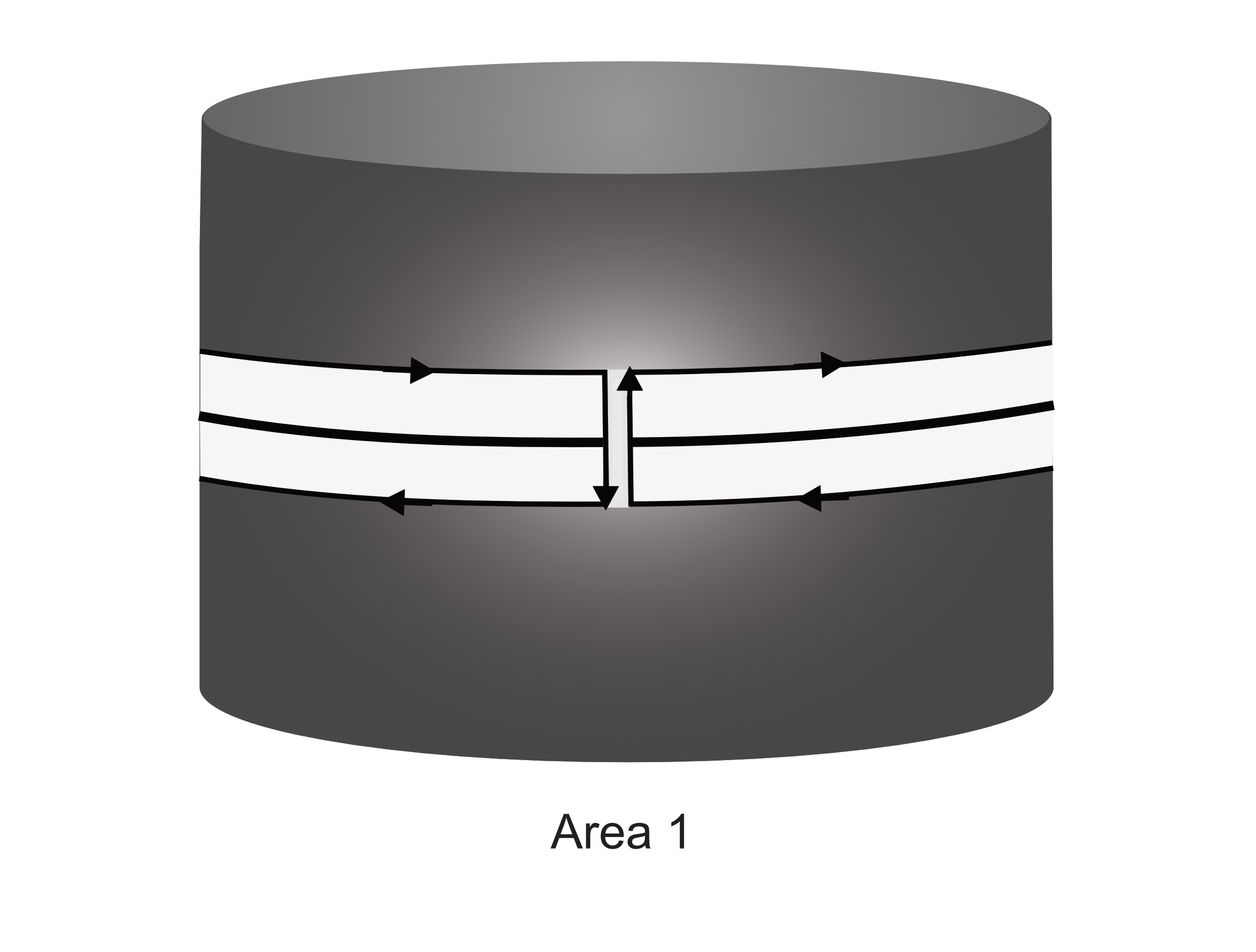}
 \caption{Area 1 is shaded in dark gray. This is one of the possible areas bounded by the line integral contour of $\oint {\bf A} \cdot d{\bf l}$ from Fig \eqref{fig1} for evaluating $\int {\bf B} \cdot d {\bf a}$.}
\label{fig2}
\end{figure}

\begin{figure}[ht]
 \centering
\includegraphics[width=100mm]{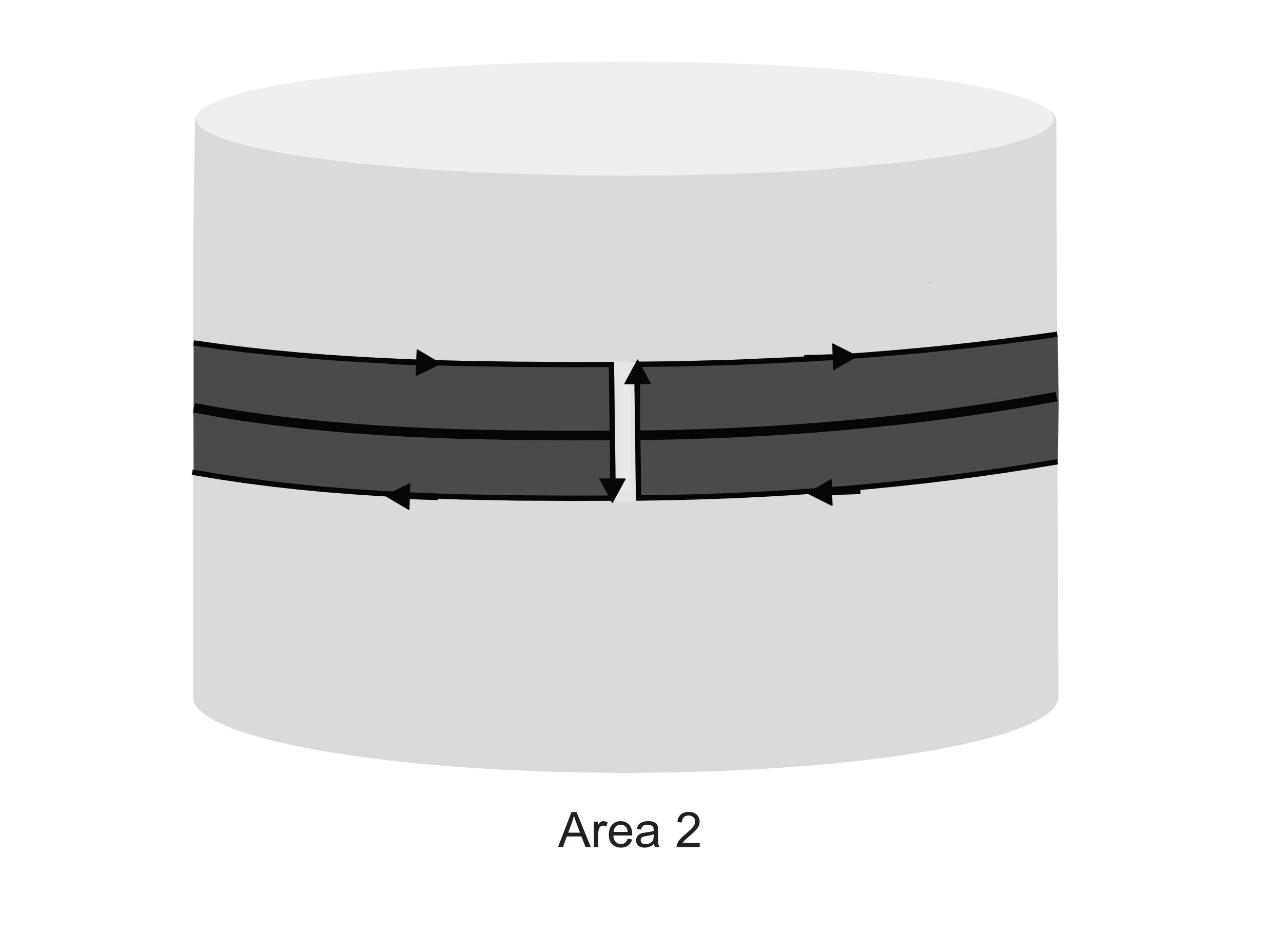}
 \caption{Area 2 is shaded in dark gray. This is the other option for the area bounded by the line integral contour of $\oint {\bf A} \cdot d{\bf l}$ from Fig \eqref{fig1} for evaluating $\int {\bf B} \cdot d {\bf a}$.}
\label{fig3}
\end{figure}

Reference \cite{siva2} gives a more direct construction for obtaining the result in \eqref{b-disk}, whereas here we show that the delta function term is required by boundary conditions and Stokes's theorem. We do this by evaluating Stokes's theorem for the magnetic field and vector potential, $\int {\bf B} \cdot d {\bf a} = \oint {\bf A} \cdot d {\bf l}$, for the line integral and two possible surfaces as shown in Figs. \eqref{fig1} \eqref{fig2} and \eqref{fig3}. 

The thin solid line with arrows in Fig. \eqref{fig1} is the contour that runs from $A \to B \to C \to D = A$. The part of the contour that is the segment $DA$ (around the back) is entirely in the northern hemisphere where the vector potential is ${\bf A}_+$; the segment $BC$ (around the back) is entirely in the southern hemisphere where the vector potential is ${\bf A}_-$; the two short vertical segments, $AB$ and $CD$ cross the equator and bridge the region where ${\bf A}_+$ and ${\bf A}_-$ overlap. The two dotted lines bracketing the equator (the thicker solid line) bound the region of width $2 \epsilon$ where the two vector potentials overlap. The solid lines of the contour above and below the equator are separated by $2 \epsilon '$.

The part of the contour $DA$ has $d{\bf l} = \rho_0 d \varphi {\bf \hat \varphi}$ and the vector potential is ${\bf A}_+$ so we have from \eqref{A-coulomb1}
\begin{equation}
\label{DA}
    \int _D ^A {\bf A}_+ \cdot d {\bf l} = \int _0 ^{2 \pi} \frac{g}{\rho _0} \left( 1- \frac{\epsilon '}{\sqrt{\rho_0 ^2 + (\epsilon')^2}}\right) {\hat \varphi} \cdot \rho_0 d \varphi {\hat \varphi}
    = 2 \pi g -\frac{2 \pi g \epsilon '}{\sqrt{\rho_0 ^2 + (\epsilon')^2}}~.
\end{equation}
For the part of the contour $AB$ that crosses the equator is in the region where the vector potentials, ${\bf A}_+$ and ${\bf A}_-$, overlap, but for this contour $d {\bf l} \propto {\bf \hat z}$ and since both ${\bf A}_\pm \propto {\bf \hat \varphi}$ and ${\bf \hat z} \cdot {\bf \hat \varphi}=0$ this part of the contour gives zero. The same is true for the part of the contour $CD$. Thus we have
\begin{equation}
\label{AB-CD}
    \int _A ^B {\bf A}_\pm \cdot d {\bf l} = \int _C ^D {\bf A}_\pm \cdot d {\bf l} =0.
\end{equation}
A more careful treatment of these parts of the contour is carried out in \cite{siva2} using the method laid out in \cite{wu-yang}, but the end result is that these parts of the contour give zero. 

Finally, the $BC$ segment has $d {\bf l} = - \rho_0 d \varphi {\bf \hat \varphi}$ since the direction of traversal is opposite to that of the segment $DA$, and here the vector potential is ${\bf A}_-$. This segment contributes to the contour as
\begin{equation}
\label{BC}
    \int _B ^C {\bf A}_- \cdot d {\bf l} = \int _0 ^{2 \pi} \frac{g}{\rho _0} \left( -1- \frac{(-\epsilon ')}{\sqrt{\rho_0 ^2 + (-\epsilon')^2}}\right) {\hat \varphi} \cdot (-\rho_0 d \varphi {\hat \varphi}) = 2 \pi g -\frac{2 \pi g \epsilon '}{\sqrt{\rho_0 ^2 + (\epsilon')^2}}~,
\end{equation}
Adding all these contributions together gives
\begin{eqnarray}
    \label{ABCD-3}
    \oint {\bf A} \cdot d{\bf l} &=& 
    \int _D ^A {\bf A}_+ \cdot d{\bf l} + \int_A ^B {\bf A} \cdot d{\bf l} + \int _B ^C {\bf A}_- \cdot d {\bf l} + \int _C ^D {\bf A} \cdot d {\bf l}   \nonumber \\
    &=& \left( 2 \pi g -\frac{2 \pi g \epsilon '}{\sqrt{\rho_0 ^2 + (\epsilon')^2}}\right) + 0 + \left(2 \pi g -\frac{2 \pi g \epsilon '}{\sqrt{\rho_0 ^2 + (\epsilon')^2}}\right) + 0 ~ \\
    &=& 4 \pi g - \frac{4 \pi g \epsilon '}{\sqrt{\rho_0 ^2 + (\epsilon')^2}} \nonumber 
\end{eqnarray}

Next, we evaluate $\int {\bf B} \cdot d {\bf a}$. From Fig. \eqref{fig1} there are two possible areas spanned by the contour as shown in Figs. \eqref{fig2} and \eqref{fig3}. The first area is the entire surface minus the strip of width $2 \epsilon '$ around the equator. For this area (labeled Area 1) the additional delta function part of the magnetic field in \eqref{b-disk} does not contribute since it lies in the thin strip that is excluded by Area 1. Thus, the magnetic flux through Area 1 is only due to the Coulomb part of \eqref{b-disk} namely ${\bf B}_{{\rm Coulomb}} = g \frac{\rho {\hat {\bf \rho}} + z {\hat {\bf z}}}{(\rho ^2 +z^2)^{3/2}}$. For the whole cylindrical surface, this flux is the standard result $\int_{{\rm full-surface}} {\bf B}_{{\rm Coulomb}} \cdot d{\bf a} = 4 \pi g$. Thus we just need to find the magnetic flux due to the Coulomb magnetic field through the thin strip and subtract this from $4 \pi g$. For this thin strip of width $2 \epsilon '$ the Coulomb magnetic field is ${\bf B}_{{\rm Coulomb}} = g \frac{\rho_0 {\hat {\bf \rho}} + z {\hat {\bf z}}}{(\rho_0 ^2 +z^2)^{3/2}}$, and the infinitesimal area is $d{\bf a} =  \rho _0 dz d \varphi {\bf \hat \rho}$, and thus the flux through the thin strip is 
\begin{equation}
    \label{strip}
    \int _{{\rm Strip}} {\bf B}_{{\rm Coulomb}} \cdot d{\bf a}  = g  \int _0 ^{2 \pi} \rho _0 d \varphi \int _{-\epsilon'} ^{\epsilon'} dz \frac{\rho_0}{(\rho_0 ^2 +z^2)^{3/2}} = 4 \pi g \frac{\epsilon'}{\sqrt{\rho_0 ^2 + (\epsilon')^2}}~,
\end{equation}
which is the missing flux due to the thin strip in Fig. \eqref{fig1}. Subtracting the result \eqref{strip} from the full cylindrical surface yields the flux through Area 1
\begin{equation}
    \label{area-1}
    \int _{{\rm Area-1}} {\bf B}_{{\rm Coulomb}} \cdot d{\bf a}  = 4 \pi g - \frac{4 \pi g\epsilon'}{\sqrt{\rho_0 ^2 + (\epsilon')^2}} = \oint {\bf A} \cdot d {\bf l}~,
\end{equation}
which agrees with the result of the line integral, $\oint {\bf A} \cdot d{\bf l}$ in equation \eqref{ABCD-3}.

Next we will evaluate $\int {\bf B} \cdot d {\bf a}$ for Area 2 in Fig. \eqref{fig3} . This should again yield the same result as \eqref{area-1} since both areas are bounded by the same contour. For Area 2 the area element has the opposite sign from \eqref{area-1}, namely $d{\bf a} =  - \rho _0 dz d \varphi {\bf \hat \rho}$, since the direction of traversal of the contour is reversed for this area. Thus for Area 2 the Coulomb magnetic field will give the same magnitude as the result in \eqref{strip} but opposite sign 
\begin{equation}
    \label{area-2}
    \int _{Area-2} {\bf B}_{Coulomb} \cdot d{\bf a}  = - 4 \pi g \frac{\epsilon'}{\sqrt{\rho_0 ^2 + (\epsilon')^2}}~.
\end{equation}
This result does not match the result for $\oint {\bf A} \cdot d{\bf l}$ from \eqref{ABCD-3} and would violate Stokes's theorem. We now show that the extra non-Coulomb piece is exactly what is needed by Stokes's theorem. 
\begin{eqnarray}
    \label{area-2a}
    \int _{{\rm Area-2}} {\bf B}_{{\rm extra}} \cdot d{\bf a}  &=&  \int _{\rm{Area-2}} \left( - \frac{2g \delta (z)}{\rho_0} {\hat {\bf \rho}}~.\right) \cdot (- \rho _0 dz d \varphi {\bf \hat \rho}) \nonumber \\
    &=& 2g \int_{-H} ^H \delta (z) dz \int_0 ^{2 \pi} d \varphi = 4 \pi g ~.
\end{eqnarray}
Combining \eqref{area-2} and \eqref{area-2a} gives
\begin{equation}
    \label{area-2b}
    \int _{Area-2} ({\bf B}_{extra} + {\bf B}_{Coulomb} )\cdot d{\bf a}  = 4 \pi g - \frac{4 \pi g\epsilon'}{\sqrt{\rho_0 ^2 + (\epsilon')^2}} = \oint {\bf A} \cdot d {\bf l}~,
\end{equation}
which again matches with the line integral in \eqref{ABCD-3}, showing that the extra, second term in \eqref{b-disk} is needed to make Stokes's theorem work out. 

To conclude this section, we calculate the current density associated with the magnetic field from \eqref{b-disk} using $\nabla \times {\bf B} = 4 \pi {\bf J}$. The Coulomb term in \eqref{b-disk} has a curl of zero, since it is radially symmetric and thus does not lead to a current density. The disk term gives
\begin{equation}
    \label{sheet}
    \nabla \times \left( - \frac{2g \delta (z)}{\rho} {\hat {\bf \rho}} \right) = -\frac{2g}{\rho} \frac{d(\delta (z))}{dz} {\hat {\bf \varphi}} = -\frac{2g}{\rho} \delta ' (z) {\hat {\bf \varphi}}~,
\end{equation}
Thus we find a current density of
\begin{equation}
\label{j3}
    {\bf J}_e = -\frac{g}{2 \pi \rho}  \delta ' (z) {\hat {\bf \varphi}} ~. 
\end{equation}

The fact that the current density is proportional to the derivative of a delta function implies that there is a {\it dipole} current density at $z=0$. The above analysis of the jump in the vector potential over a short distance and with a source that is the derivative of a delta function is essentially the magnetic version of the charge dipole layer jump of the scalar potential discussed in section 1.6 of Jackson's electrodynamics text \cite{jackson}.

\section{Field Momentum of magnetic plus electric charges}

We begin by placing the field configuration of the Dirac model \eqref{B-coulomb}, the Banderet model \eqref{band-2} or the Wu-Yang model \eqref{b-disk} of magnetic charge at the origin.  Next, we place an electric charge at some point away from the origin such as ${\bf r}_0$.  The electric field will be a Coulomb field ${\bf E} = q \frac{\bf \hat r'}{r'^2}$ with ${\bf r}' = {\bf r} - {\bf r}_0$ being the radial vector from the location of the electric charge. It is well known \cite{jackson,shnir,heras,olive,blag,felsager,adorno,mavromatos} that this system of magnetic charge plus electric charge carries a field angular momentum of ${\bf L}_{EM} = - qg {\bf \hat r}_0$ in the Coulomb electric and Coulomb magnetic fields, with ${\bf \hat r}_0$ being a unit vector pointing from the electric charge to the magnetic charge. One can use this result to obtain the Dirac condition \eqref{dirac-cond} by requiring that the magnitude of this field angular momentum be some integer multiple of the fundamental unit of angular momentum $\frac{\hbar}{2}$ {\it i.e.} $|{\bf L}_{EM}| =  qg = n \frac{\hbar}{2}$ which is the Dirac condition in \eqref{dirac-cond}. \footnote{Reference \cite{dunia} showed that the string magnetic field gives an additional field angular momentum that spoils the use of angular momentum quantization to obtain the Dirac quantization condition.} 

It is also well known and easy to show that there is {\it no} linear momentum carried by the Coulomb magnetic field and the Coulomb electric field of the monopole plus electric charge system, namely ${\bf P}_{{\rm EM}} ^{{\rm Coulomb}} = \frac{1}{4 \pi} \int q \frac{{\bf r}'}{(r')^3} \times g \frac{{\bf r}}{r^3} d^3 x = 0$. However, as we will show in the next three subsections, the non-Coulomb parts of the magnetic fields for the Dirac \eqref{B-coulomb}, Banderet \eqref{band-2} or Wu-Yang \eqref{b-disk} models {\it do} give a nonzero field momentum. 

\subsection{Field Momentum of Dirac String plus electric charge}

If we insert the non-Coulomb part of the Dirac string magnetic field from \eqref{B-coulomb}, along with the Coulomb electric field into the field momentum expression \eqref{momentum} we find
\begin{equation}
\label{mom3d}
{\bf P}_{{\rm EM}} ^{{\rm String}} = \frac{1}{4 \pi} \int  \left (q  \frac{{\bf \hat r}'}{{r'}^2} \times ( \pm 4 \pi g \delta (x) \delta (y) \Theta (\mp z) {\bf {\hat z}}) \right) d^3 x  ~.
\end{equation} 
Because of the form of the string magnetic field, it is easiest to evaluate the integrals in \eqref{mom3d} in Cartesian coordinates so that ${\bf r}' = {\bf r} - {\bf r}_0 = (x-x_0){\bf \hat x} + (y-y_0){\bf \hat y} + (z-z_0){\bf \hat z}$. The $dx$ and $dy$ integrations are simple to carry out using the delta functions $\delta (x)$ and $\delta (y)$. This leaves a simple $dz$ integration which yields \cite{siva}
\begin{equation}
\label{mom3d-1}
{\bf P}_{{\rm EM}} ^{{\rm String}} 
= \pm gq \int _{-\infty} ^{\infty} \Theta (\mp z) \frac{x_0 {\bf \hat y} - y_0 {\bf \hat x}}{(\rho_0 ^2 + (z-z_0)^2)^{3/2}} dz
= \pm gq \frac{(r_0 \mp z_0)}{r_0 \rho _0 ^2} (-y_0 {\bf \hat x} + x_0 {\bf \hat y})  ~,
\end{equation} 
with $\rho^2 _0 = x_0 ^2+ y_0^2$ and $r_0 ^2 = x_0 ^2+ y_0 ^2 + z_0 ^2 = \rho_0^2 + z_0^2$. This nonzero field momentum may initially seem unusual. If for the moment one ignores the current density in \eqref{j} then nothing in the system is moving -- the electric and magnetic charges are stationary, and the fields are static. Nevertheless, there is a non-zero linear field momentum in the system. If \eqref{mom3d-1} is the only momentum in the system, this would violate the center-of-energy theorem \cite{coleman,zangwill}. This violation of the center-of-energy theorem suggests that ignoring the current density in \eqref{j} is not a feasible option. In Section IV we do not neglect the current density \eqref{j} and show that this system carries a hidden momentum in the charges and current densities, which balances the field momentum. 

\subsection{Field Momentum of Banderet Potential plus electric charge}

Next we calculate the field momentum for the non-Coulomb part of the Banderet magnetic field  -- the second term in \eqref{band-2} -- which is 
\begin{equation}
\label{band-4}
{\bf P}_{{\rm EM}} ^{{\rm Banderet}} = -\frac{2 \pi qg}{4 \pi} \int  \Theta (x) \delta (y) x  \frac{{\bf r}'}{(r')^3} \times \frac{{\bf r}}{r^3} d^3 x  ~.
\end{equation}
We choose the electric charge to be located at ${\bf r}_0 = y_0 {\bf \hat y}$. This is not the most general case, but taking a more general location of the electric charge did not lead to an analytic result for the integral in \eqref{band-4}. With this choice of ${\bf r}_0$, we have ${\bf r}' = {\bf r} - y_0 {\bf \hat y}$ and $r' = (x^2 + (y-y_0)^2 +z^2)^{1/2}$. Using this in \eqref{band-4} and carrying out the $dy$ integration gives
\begin{eqnarray}
\label{band-5}
{\bf P}_{{\rm EM}} ^{{\rm Banderet}} &=& \frac{qgy_0}{2}  \int _0 ^\infty dx \int _{-\infty} ^{\infty} dz
\frac{(-x^2 {\bf \hat z}+ x z {\bf \hat x})}{(x^2+y_0^2 +z^2)^{3/2} (x^2+ z^2)^{3/2}}  \\
&=& - \frac{qgy_0}{2}  \int _0 ^\infty dx \int _{-\infty} ^{\infty} dz
\frac{x^2 {\bf \hat z}}{(x^2+y_0^2 +z^2)^{3/2} (x^2+ z^2)^{3/2}}~. \nonumber
\end{eqnarray}
We have taken into account the $\Theta (x)$ in the integration range of the $dx$ integration. Also in going from the first line to the second we have carried out the $dz$ integration for the ${\bf \hat x}$ component, which gives zero since the integration range is over all $z$, from $-\infty$ to $+\infty$, and the integrand, as a function of $z$, is an odd function. Finally we use {\it Mathematica} to carry out the $dx$ and $dz$ integrations in the last line of \eqref{band-5} yielding
\begin{equation}
\label{band-6}
{\bf P}_{{\rm EM}} ^{{\rm Banderet}} = - \frac{qg \pi}{4 y_0} {\bf \hat z}~.
\end{equation}
As for the Dirac string potential, we see that there is apparently an uncanceled field momentum. In Section IV we will show that this system also carries hidden momentum in the charges and currents that give rise to the full magnetic field in \eqref{band-2}.

\subsection{Field Momentum of Wu-Yang Fiber Bundle plus electric charge}

We now calculate the field momentum of an electric charge plus the disk magnetic field of the Wu-Yang monopole model, the second term in \eqref{b-disk}. By cylindrical symmetry and without loss of generality, we place the electric charge $q$ at ${\bf r}_0 = \rho _0 {\bf \hat x} + z_0 {\bf \hat z}$ giving an electric field of ${\bf E} = q \frac{{\bf r} '}{r'^3}$ where ${\bf r}' = {\bf r} - {\bf r}_0$. In cylindrical coordinates $r' = (\rho ^2+ \rho_0 ^2 - 2 \rho \rho_0 \cos \varphi+ (z-z_0)^2)^{1/2}$. With this set up the field momentum of the electric charge plus disk magnetic field is 
\begin{equation}
\label{momentum-wy}
{\bf P}_{{\rm EM}} ^{{\rm disk}} = \frac{1}{4 \pi} \int q  \frac{{\bf \hat r}'}{{r'}^2} \times \left( - \frac{2 g \delta (z)}{\rho} {\bf {\hat \rho}} \right) d^3 x
\end{equation} 
Using $\delta (z)$ to handle the $dz$ integration gives
\begin{eqnarray}
\label{momentum-wy2}
{\bf P}_{{\rm EM}} ^{{\rm disk}} 
&=& - \frac{qg}{2 \pi} \int _0 ^\infty \rho d \rho \int _0 ^{2 \pi} d \varphi  \frac{\left[ \rho {\bf \hat \rho} - \rho_0 {\bf \hat x} -z_0 {\bf \hat z} \right]}{(\rho ^2 + \rho_0 ^2 + z_0 ^2 -2 \rho \rho_0 \cos \varphi)^{3/2}}\times \frac{{\bf \hat \rho}}{\rho}  \\
&=&-\frac{qg}{2 \pi} \int _0 ^\infty  d \rho \int _0 ^{2 \pi} d \varphi  \left( \frac{z_0 \sin \varphi {\bf \hat x} - z_0 \cos \varphi {\bf \hat y} -\rho_0 \sin \varphi {\bf \hat z} }{(\rho ^2 + \rho_0 ^2 + z_0 ^2 -2 \rho \rho_0 \cos \varphi)^{3/2}} \right) \nonumber ~.
\end{eqnarray} 
In going from the first line to the second we have carried out the cross product. The ${\bf \hat x}$ and ${\bf \hat z}$ components in the last line of \eqref{momentum-wy2} are of the same form, and carrying out the $d \varphi$ integration gives zero for both. This leaves the field momentum as
\begin{equation}
\label{momentum-wy3}
{\bf P}_{{\rm EM}} ^{{\rm disk}} =
\frac{qgz_0}{2 \pi} \int _0 ^\infty  d \rho \int _0 ^{2 \pi} d \varphi \frac{\cos \varphi {\bf \hat y}}{(\rho ^2 + \rho_0 ^2 + z_0 ^2 -2 \rho \rho_0 \cos \varphi)^{3/2}} ~.
\end{equation} 
Using {\it Mathamatica} the integration of the ${\bf \hat y}$ component gives
\begin{equation}
    \label{momentum-2}
    {\bf P}_{{\rm EM}} ^{{\rm disk}} = - \frac{qg z_0}{\rho _0 r_0} {\bf \hat y}
\end{equation}
${\bf P}_{{\rm EM}} ^{{\rm disk}} \ne 0$ except for two special cases when $z_0=0$  for $q$ in the $z$-plane, and $\rho_0 = 0$ for $q$ along the $z$-axis. In both cases ${\bf P}_{{\rm EM}} ^{{\rm disk}} = 0$. For the case $\rho_0 =0$, the $\varphi$ integral in \eqref{momentum-wy3} is zero, and the $\rho$ integral is finite. 

\section{Hidden Mechanical Momentum}

In the above sections IIIA, IIIB, IIIC we found that the Dirac, Banderet, and Wu-Yang models all have a non-zero field momentum in the electric and magnetic fields when an electric charge, $q$, is placed in the vicinity of the magnetic fields in equations \eqref{B-coulomb}, \eqref{band-2}, and \eqref{b-disk}. As previously pointed out, this seems problematic vis-a-vis the center-of-energy theorem \cite{coleman,zangwill} since this system has a net momentum, and yet both charges are stationary and the fields are static. Such systems have been analyzed before \cite{shockley} and have led to apparent paradoxes/problems, which were resolved \cite{coleman} via ``hidden" mechanical momentum {\it i.e.} momentum carried by the currents and charges that make up the system.  Below we show that each of these three monopole models -- Dirac, Banderet, and Wu-Yang -- carries such a hidden momentum, which resolves the issue with the center-of-energy theorem. 

Good discussions of ``hidden" mechanical momentum can be found in references \cite{griffiths,griffiths-2}. A particularly transparent example of ``hidden" momentum is given in \cite{griffiths-text} \cite{ph} which involves a rectangular current loop of moving charges embedded in a uniform electric field. In this rectangle loop example, the hidden momentum comes from the $\gamma$ factor in the relativistic expression for momentum. The appearance of the $\gamma$ factor shows that hidden momentum can occur in systems where there are no electromagnetic fields at all. One can generalize the specific example given in \cite{griffiths-text} and \cite{ph} to give an expression for the ``hidden" mechanical momentum stored in a system of steady currents and static electric field:  
 \begin{equation}
     \label{hid-mom}
     {\bf P} _{{\rm hid}} = - \int \phi _e {\bf J}_e d^3x~,
 \end{equation}
where $\phi_e$ is the scalar potential of the charges and ${\bf J}_e$ is the current density. In our case $\phi_e$ is the scalar potential of $q$ and ${\bf J}_e$ is the current density of each monopole model given in equations \eqref{j}, \eqref{j2} and \eqref{j3}.

One subtle point about the expression for hidden mechanical momentum given in \eqref{hid-mom} is that it could be argued to be circular since for the field's linear momentum one can write 
 \begin{eqnarray}
     \label{hid-mom-1}
     {\bf P}_{{\rm EM}} &=& \frac{1}{4 \pi } \int ({\bf E} \times {\bf B}) d^3 x = -\frac{1}{4 \pi } \int (\nabla \phi _e \times {\bf B}) d^3 x \nonumber \\
     &=& -\frac{1}{4 \pi} \int  \nabla \times (\phi_e {\bf B}) d^3 x + \frac{1}{4 \pi} \int \phi_e \nabla  \times {\bf B} d^3 x \\
     &=& -\frac{1}{4 \pi} \oint d{\bf a} \times (\phi_e {\bf B} )+ \int \phi_e {\bf J}_e d^3 x \nonumber
 \end{eqnarray}
In arriving at the result in the last line we have used Stokes' theorem to obtain the first term and used $\nabla \times {\bf B} = 4 \pi {\bf J}_e$ for the second term. Usually, the first surface term in \eqref{hid-mom-1} vanishes for localized sources and fields, by taking the surface to infinity. However, while the point electric source that produces $\phi_e$ does fall off with distance, the string part of ${\bf B}$ (see the second term in \eqref{B-coulomb}) extends to infinity. This calls into question the ability to drop the surface term in \eqref{hid-mom-1}. If one drops the surface term in \eqref{hid-mom-1}, then one sees in general that ${\bf P} _{{\rm hid}} =- {\bf P}_{{\rm EM}}$ if \eqref{hid-mom} is used to calculate the hidden momentum. One might argue for dropping the surface term in \eqref{hid-mom-1}, despite the string part of ${\bf B}$ extending to infinity, since the string contribution has zero extent due to the $\delta (x) \delta (y)$ term. Alternatively, one could argue for dropping the surface term if the surface is taken to be spherical, since for a spherical surface $d{\bf a} \propto {\hat {\bf r}}$ and thus $d {\bf a} \times {\bf B} = 0$ along the $z$-axis of the spherical surface. However, dropping the surface term is not as straightforward as is generally the case for localized fields and sources. For completeness, we present the explicit calculations confirming the cancellation of the electromagnetic field momentum of the three monopole models with the hidden mechanical momentum in Appendix C. 

\section{Summary and Conclusions}

In this paper, we examined three models for magnetic charge: the Dirac string potential \cite{dirac,dirac1}, the Banderet potential \cite{banderet}, and the Wu-Yang fiber bundle approach \cite{wu-yang,yang}. For each of these three models there is some non-Coulomb magnetic field. For the Dirac approach, the Dirac string is given by the second term in \eqref{B-coulomb}. For the Banderet approach there is a semi-infinite disk magnetic field given by the second term in \eqref{band-2}. For the Wu-Yang approach there is the disk magnetic field in the $xy$-plane given by the second term in \eqref{b-disk}. 

We have shown that these non-Coulomb magnetic fields plus the Coulomb electric field of $q$ give rise to a non-zero field momentum despite the magnetic charge $g$, and the electric charge, $q$, being at rest. By itself, this uncanceled field momentum -- given in equations \eqref{mom3d-1}, \eqref{band-6} and \eqref{momentum-2} for the Dirac, Banderet, and Wu-Yang models, respectively -- would violate the center-of-energy theorem \cite{coleman,zangwill}. However, associated with each of these extra, non-Coulomb magnetic fields there was a current density ${\bf J}_e$, given in equations \eqref{j} \eqref{j2} and \eqref{j3} for the Dirac, Banderet and Wu-Yang cases, respectively. These current densities combined with the scalar potential, $\phi$, of the electric charge $q$ to give each of these systems a hidden mechanical momentum $ {\bf P}^{{\rm mech}} _{{\rm hid}} = - \int \phi_e {\bf J} d^3x$. Calculating the hidden mechanical momentum for the three magnetic charge models -- the results are given in \eqref{p-dirac-2} \eqref{p-band2} and \eqref{dphi} for the Dirac, Banderet and Wu-Yang cases, respectively -- showed that the field momentum plus hidden mechanical momentum cancel each other out for the three magnetic charge models {\it i.e.} ${\bf P}_{{\rm field}} + {\bf P}^{{\rm hid}} _{{\rm mech}} =0$. This saves the center-of-energy theorem and reinforces the fact that the non-Coulomb magnetic fields of \eqref{B-coulomb}, \eqref{band-2} and \eqref{b-disk} and their associated current densities of \eqref{j} \eqref{j2} and \eqref{j3} are needed for the internal consistency of these models. Previous works have argued for the physical reality of features such as the Dirac string based on their gravitational effects \cite{banyas}.

Although the above analysis calls into question the standard, Abelian monopole models based on a single Abelian vector potential, Cabibbo-Ferrari \cite{cabibbo,zwanziger,singleton-1996} models of magnetic charge with two vector potentials are still viable, as are non-Abelian 't Hooft-Polyakov monopole models \cite{thooft,polyakov} or recently investigated electroweak monopoles \cite{hung1,hung2}.  

\appendix

\section{Current Density for Banderet potential}
In this appendix we calculate the current density for the Banderet potential. The current density is given by ${\bf J}_e = \frac{\nabla \times {\bf B}}{4 \pi}$. Using the magnetic field from \eqref{band-2} gives
\begin{equation}
    \label{ap1}
    {\bf J} = -\frac{g}{2} \nabla \times \left( \Theta (x) \delta (y) x \frac{{\bf \hat r}}{r^2}\right) = \frac{g}{2} \frac{\hat {\bf r}}{r^2} \times \left( \delta (y) \delta (x) x {\bf \hat x} + \Theta (x) x \delta ' (y) {\bf \hat y} + \Theta (x) \delta (y) {\bf \hat x} \right) ~.
\end{equation}
In \eqref{ap1}  we have used the product rule $\nabla \times (f{\bf A}) = f \nabla \times {\bf A} -  {\bf A} \times \nabla (f) $, with ${\bf A} = \frac{{\bf \hat r}}{r^2}$ and $f=  \Theta (x) \delta (y) x$. Also $\nabla \times \frac{{\bf \hat r}}{r^2} =0 $ and the primes in \eqref{ap1} are derivatives with respect to the argument of the function. Now, the first term in \eqref{ap1} is zero since $x \delta (x)$ is zero. Next using ${\bf \hat r} = \frac{\bf r}{r}$ and ${\bf r} = x {\bf \hat x} + y {\bf \hat y} + z {\bf \hat z}$ and ${\bf r} \times {\bf \hat y} = x {\bf \hat z} - z {\bf \hat x}$ and ${\bf r} \times {\bf \hat x} = - y {\bf \hat z} + z {\bf \hat y}$ the current density becomes
\begin{eqnarray}
    \label{ap2}
    {\bf J}_e &=& \frac{g\Theta (x)}{2 r^3}  \left( x^2 \delta ' (y) {\bf \hat z}  - xz \delta ' (y) {\bf \hat x} - y \delta (y) {\bf \hat z} +  z \delta (y) {\bf \hat y}\right) \nonumber \\
    &\to&  \frac{g\Theta (x)}{2 r^3}  \left( x^2 \delta ' (y) {\bf \hat z}  -xz \delta ' (y) {\bf \hat x}  +  z \delta (y) {\bf \hat y}\right)~,
\end{eqnarray}
where in the last term we have used $y \delta (y) \to 0$.

\section{Magnetic flux of delta function terms}

In this appendix, we show that the magnetic flux of all three monopole models -- Dirac, Banderet, and Wu-Yang -- give an inward flux of $4 \pi g$ to match the outward flux of $4 \pi g$ due to the Coulomb part of the magnetic field. 

First for the Coulomb part of the magnetic field ({\it i.e.} ${\bf B}_{{\rm Coulomb}} = \frac{g {\bf \hat r}}{r^2}$) the flux through a sphere of radius $R$ is given by 
\begin{equation}
    \label{b-flux}
    \oint {\bf B}_{{\rm Coulomb}} \cdot d {\bf a} = \int g R^2   \frac{ {\bf \hat r}}{R^2} \cdot {\bf \hat r} d \Omega= 4 \pi g ~,
\end{equation}
since the integral over the full solid angle is $4 \pi$.

For the Dirac version of the monopole, the string part of the magnetic field is $\pm 4 \pi g \delta (x) \delta (y) \Theta (\mp z) {\bf \hat z}$ -- see \eqref{B-coulomb}. The $\delta (x) \delta (y)$ term fixes the strings for the two potentials to run along the $z$-axis. For the string along the $-z$ axis the magnetic field is $+ 4 \pi g \delta (x) \delta (y) \Theta (- z) {\bf \hat z}$ and $d {\bf a} = - dx dy {\bf \hat z}$ giving
\begin{equation}
    \label{b-flux2}
    \oint {\bf B}_- ^{{\rm string}} \cdot d {\bf a} = \int 4 \pi g \delta (x) \delta (y){\bf \hat z} \cdot (- {\bf \hat z} dx dy ) = - 4 \pi g ~,
\end{equation}
For the string along the $+z$ axis the magnetic field is   $- 4 \pi g \delta (x) \delta (y) \Theta (+ z) {\bf \hat z}$ and $d {\bf a} = dx dy {\bf \hat z}$ which gives
\begin{equation}
    \label{b-flux3}
    \oint {\bf B}_+ ^{{\rm string}} \cdot d {\bf a} = \int - 4 \pi g \delta (x) \delta (y){\bf \hat z} \cdot ({\bf \hat z} dx dy ) = - 4 \pi g ~.
\end{equation}
In both cases the string contribution gives a flux of $-4 \pi g$ to balance the flux of $+ 4 \pi g$ coming from the Coulomb part {\it i.e.} $\Phi_{{\rm Coulomb}} + \Phi_{{\rm string}} = 4 \pi g - 4\pi g =0$. This is the flux through any closed surface containing the origin. 

For the Banderet version of the monopole, the extra magnetic field given in \eqref{band-2} is $ - 2 \pi g \Theta (x) \delta (y) x \frac{{\bf \hat r}}{r^2}$. We use a spherical surface of radius $R$ to calculate the magnetic flux of this extra piece which gives
\begin{eqnarray}
\label{flux-band}
\oint {\bf B}^{{\rm band}} \cdot d{\bf a} &=& -2 \pi g \int \Theta (x) \delta (y) x \frac{{\bf \hat r}}{R^2} \cdot {\bf \hat r} R^2 \sin \theta d \theta d \varphi \\
&=& -2 \pi g \int \Theta (R \sin \theta \cos \varphi) \delta (R \sin \theta \sin \varphi ) R \sin ^2 \theta \cos \varphi d \theta d \varphi \nonumber ~.
\end{eqnarray}
In the last step, we have converted everything to spherical polar coordinates. Now we carry out the $d \varphi$ integration and use the delta function property $\delta (f(x)) = \sum _i \frac{1}{|f'(x)|_{x=x_i}} \delta(x-x_i)$, where $x_i$ are the zeros of $f(x)$ and $f'(x)$ is the derivative of $f(x)$. Now for the delta function in \eqref{flux-band} $f'(x) \to R \sin \theta \cos \varphi$ and the zeros occur at $\varphi =0$ and $\varphi = \pi$. However, because of the $\Theta (R \sin \theta \cos \varphi)$ term in \eqref{flux-band} only $\varphi =0$ is in the range of integration since $\Theta (R \sin \theta \cos \varphi) =0$ when $\varphi = \pi$. With all this the $\varphi$ integration is
\begin{equation}
    \label{flux-band-2}
\int _0 ^{2 \pi} R \sin ^2 \theta \cos \varphi \frac{1}{R \sin \theta \cos \varphi} \delta(\varphi -0) d \varphi = \sin \theta 
\end{equation}
We now finish by evaluating the $d \theta$ integration in \eqref{flux-band} which gives 
\begin{equation}
\label{flux-band-3}
\oint {\bf B}^{{\rm band}} \cdot d{\bf a} = -2 \pi g \int _0 ^\pi \sin \theta d \theta  = -4 \pi g ~.
\end{equation}
Thus, the Coulomb magnetic flux and Banderet magnetic flux balance {\it i.e.} $\Phi_{{\rm Coulomb}} + \Phi_{{\rm band}} = 4 \pi g - 4\pi g =0$.

For the Wu-Yang version of the monopole, the extra magnetic field given in \eqref{b-disk} is $- \frac{2g \delta (z)}{\rho} {\hat {\bf \rho}}$. We use a cylindrical surface of radius $R$ and height $2H$ going from $z=-H$ to $z=H$. The magnetic flux due to the extra disk term is
\begin{eqnarray}
\label{flux-wu-yang}
\oint {\bf B}^{{\rm disk}} \cdot d{\bf a} = -2 g \int \delta (z) \frac{{\bf \hat \rho}}{R} \cdot {\bf \hat \rho} R  d z d \varphi
= -2  g \int_0 ^{2 \pi} \int _{-H} ^H \delta (z)  dz d \varphi = -4 \pi g ~,
\end{eqnarray}
where the $dz$ integration gives $1$ by the property of the delta functions and the $d\varphi$ integration gives $2 \pi$. The end caps of the cylindrical surface contribute nothing, because ${\bf B} \perp d{\bf a}$. Thus, the Coulomb magnetic flux and disk magnetic flux balance {\it i.e.} $\Phi_{{\rm Coulomb}} + \Phi_{{\rm disk}} = 4 \pi g - 4\pi g =0$.

\section{Hidden Mechanical Momentum for three magnetic charge models}

This appendix presents explicit calculations for the hidden mechanical momentum for the Dirac, Banderet, and Wu-Yang models for magnetic charge.

\subsection{Hidden Mechanical Momentum of Dirac String}

To calculate the hidden momentum for the Dirac string case we need the scalar potential, $\phi$, for the electric charge and the current density for the Dirac string. The current density was calculated in section IIA and the result is given in equation \eqref{j} as ${\bf J}_e = \pm g \Theta (\mp z) \left[ \delta (x) \delta ' (y) {\bf \hat x} - \delta (y) \delta' (x) {\bf \hat y} \right]$. Due to the cylindrical symmetry of the Dirac string potential we can place, without loss of generality, the electric charge anywhere in the $xy$-plane on a ring of radius $\rho_0^2 = x^2_0 + y_0^2$. Thus we will choose the electric charge, $q$, to be on the $x$-axis at $x_0 = \rho_0$ and along the $z$-direction at $z=z_0$. This leaves $y_0=0$. With these choices the scalar potential becomes
\begin{equation}
    \label{pot-dirac}
    \phi_e = \frac{q}{\sqrt{(x-x_0)^2+y^2+(z-z_0)^2}} ~.
\end{equation}
With this setup the hidden mechanical momentum for the Dirac string case, with the string along the negative $z$-axis, becomes
 \begin{equation}
     \label{p-dirac}
     {\bf P}_{{\rm hid}} ^{{\rm String}} = - gq \int_{-\infty} ^0 dz \int _{-\infty} ^{\infty} dy \int _{-\infty} ^{\infty} dx
     \frac{\left[ \delta (x) \delta ' (y) {\bf \hat x} - \delta (y) \delta' (x) {\bf \hat y} \right]}{\sqrt{(x-x_0)^2 +y^2+(z-z_0)^2}}  ~.
 \end{equation}
The $\Theta (- z)$ function in the current density has been taken into account via the limits in the $dz$ integration. The case with the string along the positive $z$-axis works out identically except for some changes in signs. To handle the $\delta ' (y)$ and $\delta ' (x)$ terms we use integration by parts. For the $\delta ' (y)$ term the $dy$ integration by parts gives 
\begin{equation}
    \label{y-int-parts}
    \frac{\delta (y)}{\sqrt{(x-x_0)^2 +y^2+z^2}} {\bigg \vert}_{-\infty} ^{~\infty} + \int _{-\infty} ^{\infty} \frac{y \delta (y)}{((x-x_0)^2 +y^2+(z-z_0)^2)^{3/2}} dy = 0 ~.
\end{equation}
Both terms in \eqref{y-int-parts} are zero since $\delta (\pm \infty ) =0$, and thus the ${\bf \hat x}$ term in \eqref{p-dirac} is zero. For the $\delta ' (x)$ term the $dx$ integration by parts gives
\begin{eqnarray}
    \label{x-int-parts}
   &&\frac{\delta (x)}{\sqrt{(x-x_0)^2+y^2 +(z-z_0)^2}} {\bigg \vert}_{-\infty} ^{~\infty} + \int _{-\infty} ^{\infty} \frac{(x-x_0) \delta (x)}{((x-x_0)^2 + y^2 +(z-z_0)^2)^{3/2}} dx \nonumber \\
   &=& -\frac{x_0}{(x_0^2 + y^2+ (z-z_0)^2)^{3/2}} ~.
\end{eqnarray}
The first term in \eqref{x-int-parts} is zero since $\delta (\pm \infty) = 0$, but the second term now gives a non-zero part. Having carried out the $dx$-integration for the second term in \eqref{p-dirac}, we can immediately carry out the $dy$-integration with the help of $\delta (y)$, leaving only the $dz$ integration 
 \begin{eqnarray}
     \label{p-dirac-2}
     {\bf P}^{{\rm String}} _{{\rm hid}} &=& - gq x_0 {\bf \hat y}\int_{-\infty} ^0  
     \frac{dz}{(x_0^2 + (z-z_0)^2)^{3/2}} = - gq x_0 {\bf \hat y} \left( \frac{1}{x_0 ^2} - \frac{z_0}{x_0^2 \sqrt{x_0^2 + z_0^2}}\right)\nonumber \\
     &=& -gq\frac{(r_0 - z_0)}{r_0 x_0} {\bf \hat y} \to -gq\frac{(r_0 - z_0)}{r_0 \rho_0} {\bf \hat y} ~.
 \end{eqnarray}
 In \eqref{p-dirac-2} we have taken into account that $x_0 \to \rho_0$ since we choose $y_0=0$, and also $r_0 = \sqrt{x_0^2 +z_0^2} \to \sqrt{\rho_0^2 +z_0^2}$. For the case where the string is along the positive $z$-axis the results are similar to \eqref{p-dirac-2} but with some changes in sign giving $ {\bf P}^{{\rm String}} _{{\rm hid}} = - gq\frac{(r_0 + z_0)}{r_0 \rho_0} {\bf \hat y}$

In the expression for the field momentum from \eqref{mom3d-1} if we let $x_0 \to \rho _0$ and $y_0 \to 0$ we get ${\bf P}^{{\rm String}} _{{\rm EM}} = gq\frac{(r_0 \mp z_0)}{r_0 \rho_0} {\bf \hat y}$. Combining this field momentum with the hidden string momentum from \eqref{p-dirac-2} gives ${\bf P}^{{\rm String}} _{{\rm EM}} + {\bf P}^{{\rm String}} _{{\rm hid}} =  0$ thus preserving the center of energy theorem. 

\subsection{Hidden Mechanical Momentum of Banderet Potential}

For the field momentum calculation for the Banderet potential given in section IIIB we placed the electric charge, $q$, at ${\bf r}_0 = y_0 {\bf \hat y}$ in order to get a closed-form analytical result. This gives a scalar potential of
\begin{equation}
    \label{pot-banderet}
    \phi_e = \frac{q}{\sqrt{x^2+(y-y_0)^2+z^2}} ~.
\end{equation}
The current density for this case was calculated in equation \eqref{j2} and the result is given by ${\bf J} =   \frac{g\Theta (x)}{2 r^3}  \left( x^2 \delta ' (y) {\bf \hat z}  -xz \delta ' (y) {\bf \hat x}  +  z \delta (y) {\bf \hat y}\right)$. The hidden momentum for the Banderet model is thus
\begin{equation}
    \label{p-hid-band}
    {\bf P}_{{\rm hid}} ^{{\rm Banderet}} = - \frac{qg}{2} \int  \frac{\Theta (x)}{r^3 \sqrt{x^2+(y-y_0)^2+z^2}}\left( x^2 \delta ' (y) {\bf \hat z}  -xz \delta ' (y) {\bf \hat x}  +  z \delta (y) {\bf \hat y}\right) dx dy dz
\end{equation}

The $dz$ integration for ${\bf \hat x}$ and ${\bf \hat y}$ components of ${\bf P}_{{\rm hid}} ^{{\rm Banderet}}$ are of the form
\begin{equation}
    \label{dz-band}
    \int _{-\infty} ^{\infty} \frac{z}{(x^2+(y-y_0)^2+z^2 )^{1/2} (x^2+y^2+z^2)^{3/2}} dz = 0
\end{equation}
Since the integrand is odd and the integration range is over all of $z$, symmetry gives zero for the integration. Thus, ${\bf P}_{{\rm hid}} ^{{\rm Banderet}}$ has no ${\bf \hat x}$ or ${\bf \hat y}$ components. 

Moving on to the ${\bf \hat z}$ component of ${\bf J}_e$ and ${\bf P}_{{\rm hid}} ^{{\rm Banderet}}$, we evaluate the $dy$ integration via integration by parts
\begin{eqnarray}
    \label{dy-band}
    &&\int _{-\infty} ^{\infty} \frac{\delta ' (y) ~ dy}{(x^2+(y-y_0)^2+z^2 )^{1/2} (x^2+y^2+z^2)^{3/2}} 
    = \frac{\delta (y)}{(x^2+(y-y_0)^2+z^2 )^{1/2} (x^2+y^2+z^2)^{3/2}} \bigg{\vert} _{-\infty} ^{\infty} \nonumber \\
    &-& \int _{-\infty} ^{\infty}  \left( \frac{-3y\delta (y) ~ dy}{(x^2+(y-y_0)^2+z^2 )^{1/2} (x^2+y^2+z^2)^{5/2}} - \frac{\delta (y) (y-y_0) ~ dy}{(x^2+(y-y_0)^2+z^2 )^{3/2} (x^2+y^2+z^2)^{3/2}}\right) \nonumber \\
    &=& -\frac{y_0}{(x^2+y_0^2+z^2 )^{3/2} (x^2+z^2)^{3/2}}~.
\end{eqnarray}
The first two terms on the right-hand side of \eqref{dy-band} are zero by the property $\delta (\pm \infty) =0$.

Using the result of \eqref{dy-band} the hidden momentum for the Banderet potential is
\begin{equation}
\label{p-band}
    {\bf P}_{hid} ^{Banderet} = \frac{gqy_0 {\bf \hat z}}{2} \int _0 ^\infty x^2 dx \int ^\infty _{-\infty} \frac{dz}{(x^2+z^2)^{3/2} (x^2+y_0^2+z^2)^{3/2}}~.
\end{equation}
In \eqref{p-band} we have taken into account the $\Theta(x)$ term in ${\bf J}$ in the range of the $dx$ integration. The $dx$ integration can be done via {\it Mathematica} and yields
\begin{equation}
    \label{band-x}
    \int _0 ^\infty \frac{x^2}{(x^2+z^2)^{3/2} (x^2+y_0^2+z^2)^{3/2}} dx = \frac{-2 z^2 E(-y_0^2/z^2 ) + (y_0^2 + 2 z^2) K(-y_0^2/z^2)}{y_0^4 |z|} ~,
\end{equation}
where $K(x)$ and $E(x)$ are complete elliptic integrals of the first and second kind.  Using {\it Mathematica} to carry out the $dz$ integration of \eqref{band-x} yields
\begin{equation} 
    \label{band-z}
    \int _{-\infty} ^\infty \frac{-2 z^2 E(-y_0^2/z^2 ) + (y_0^2 + 2 z^2) K(-y_0^2/z^2)}{y_0^4 |z|} dz = \frac{\pi}{2 y_0 ^2}~.
\end{equation}
Inserting the result of \eqref{band-x} and \eqref{band-z} into \eqref{p-band} gives a hidden momentum for the Banderet model of the form
\begin{equation}
\label{p-band2}
    {\bf P}_{{\rm hid}} ^{{\rm Banderet}} = \frac{gq \pi}{4y_0} {\bf \hat z}~.
\end{equation}
Combining this hidden momentum of the Banderet model in \eqref{p-band2} with the Banderet field momentum result of \eqref{band-6} we find ${\bf P}_{{\rm EM}} ^{{\rm Banderet}} + {\bf P}_{{\rm hid}} ^{{\rm Banderet}} = 0$ and thus the center-of-energy theorem is preserved.

\subsection{Hidden Mechanical Momentum of Wu-Yang Fiber Bundle}

For the Wu-Yang fiber bundle approach, the current density is given in \eqref{j3} and has the form ${\bf J}_e = -\frac{g}{2 \pi \rho}  \delta ' (z) {\hat {\bf \varphi}}$. As in the case of the Dirac string,  the cylindrical symmetry of the disk singularity of the Wu-Yang case allows one (without loss of generality) to place the electric charge, $q$, anywhere in the $xy$-plane at a fixed radius of $\rho_0$. We again choose $q$ to be on the $x$-axis at $x=\rho_0$ and at a position of $z=z_0$ along the $z$-direction. The electric potential for $q$ is thus the same as for the Dirac string case in \eqref{pot-dirac} namely $\phi_e = \frac{q}{\sqrt{(x-\rho_0)^2+y^2+(z-z_0)^2}}$. Combining these results for the current density and scalar potential, and writing things in cylindrical coordinates the hidden momentum for the disk in the Wu-Yang set-up gives
 \begin{equation}
     \label{p-wu-yang}
     {\bf P}^{{\rm Disk}} _{{\rm hid}} = \frac{gq}{2 \pi} \int  \frac{\delta '(z) (-\sin \varphi {\bf \hat x} + \cos \varphi {\bf \hat y})}{\rho \sqrt{\rho ^2+ \rho_0^2 +(z-z_0)^2 -2 \rho_0 \rho \cos \varphi }} \rho d \rho  d \varphi dz~.
 \end{equation}
 For the ${\bf \hat x}$ component in \eqref{p-wu-yang} the $d \varphi$ integration is of the form $\int _0 ^{2 \pi} \frac{\sin \varphi }{\sqrt{A-B\cos \varphi}} d \varphi$ with $A \equiv \rho ^2+ \rho_0^2 +(z-z_0)^2$ and $B \equiv 2 \rho_0 \rho$. Using {\it Mathematica} this integration is zero, thus eliminating the ${\bf \hat x}$ component in ${\bf P}^{{\rm Disk}} _{{\rm hid}}$. 
 
 Moving on to the ${\bf \hat y}$ component in \eqref{p-wu-yang} we first carry out the $dz$ integration using integration by parts yielding
  \begin{eqnarray}
 \label{dz}
 &&\frac{gq}{2 \pi} \frac{\cos \varphi}{\rho} {\bf \hat y} \int ^\infty _{-\infty} \frac{\delta '(z)}{\sqrt{K +(z-z_0)^2}} dz \nonumber \\
 &=& \frac{gq}{2 \pi} \frac{\cos \varphi}{\rho} {\bf \hat y}\left[ \frac{\delta (z)}{\sqrt{K +(z-z_0)^2}}{\bigg \vert}_{-\infty} ^{~\infty} + \int ^\infty _{-\infty} \frac{\delta (z) (z-z_0)}{(K+(z-z_0)^2)^{3/2}}\right] \\
 &=& \frac{gq}{2 \pi} \frac{\cos \varphi}{\rho} \left(\frac{-z_0}{(K+ z_0^2)^{3/2}} \right) {\bf \hat y}~. \nonumber
\end{eqnarray}
 In \eqref{dz} we have defined $K \equiv \rho ^2+ \rho_0^2 -2 \rho_0 \rho \cos \varphi$. This first surface term in the middle line of \eqref{dz} is zero by the property $\delta (\pm \infty) =0$. We now carry out the $d \rho$ integration via {\it Mathematica} which yields
   \begin{eqnarray}
 \label{dp}
 &-&\frac{gq z_0 \cos \varphi}{2 \pi} {\bf \hat y} \int ^\infty _0 \frac{1}{\rho (\rho ^2 +\rho_0^2 + z_0^2 -2 \rho_0 \rho \cos \varphi )^{3/2}} \rho d\rho  \nonumber \\
&=& - \frac{gq z_0 \cos \varphi}{2 \pi} {\bf \hat y} \left( \frac{2(1+ \rho_0 \cos \varphi /r_0 )}{\rho_0 ^2 + 2 z_0 ^2 - \rho_0^2 \cos 2 \varphi}\right) ~,
\end{eqnarray}
Finally we carry out the $d \varphi$ integration using {\it Mathematica} yielding
\begin{eqnarray}
 \label{dphi}
{\bf P}^{{\rm Disk}} _{{\rm hid}} &=& -\frac{gq z_0 }{2 \pi} {\bf \hat y} \int ^{2 \pi} _0   \left( \frac{2\cos \varphi (1+ \rho_0 \cos \varphi /r_0 )}{\rho_0 ^2 + 2 z_0 ^2 - \rho_0^2 \cos 2 \varphi}\right) d\varphi \nonumber \\
&=&  -\frac{gq z_0 }{2 \pi} {\bf \hat y} \left( \frac{-2 \pi}{\rho_0 r_0}\right) = \frac{qg z_0}{\rho_0 r_0} {\bf \hat y}~.
\end{eqnarray}
Adding the result in \eqref{dphi} for the hidden mechanical momentum to the field momentum in \eqref{momentum-2} one finds ${\bf P}^{{\rm hid}} _{{\rm mech}} + {\bf P}^{{\rm Disk}} _{{\rm EM}} =0$ and thus the center of energy theorem is satisfied.  \\

{\bf Author Contributions:} IR and DS contributed equally to the conceptualization, formal analysis, investigation, and writing for this project. \\

{\bf Conflicts of Interest:} The authors declare no conflicts of interest.

\end{document}